\documentclass[a4paper]{article}

\usepackage[english]{babel}
\usepackage[square,sort,comma,numbers]{natbib}
\usepackage{siunitx}

\usepackage[a4paper,top=2cm,bottom=2cm,left=3cm,right=3cm,marginparwidth=1.75cm]{geometry}

\usepackage{amsmath}
\usepackage{graphicx}
\usepackage[colorlinks=true, allcolors=blue]{hyperref}

\usepackage{authblk}

\title{Visualising Ventilation Changes following Endobronchial Valve Placement with X-ray Velocimetry Functional Lung Imaging}

\author[1,2,3,*]{Ronan Smith}
\affil[1]{Adelaide Medical School, University of Adelaide, North Terrace, Adelaide, Australia}
\affil[2]{Respiratory and Sleep Medicine, Women's and Children's Hospital, King William Road, North Adelaide, Australia}
\affil[3]{Robinson Research Institute, University of Adelaide, Adelaide, Australia}

\author[4]{Charlotte Thomas}
\affil[4]{4DMedical, Melbourne, Australia}

\author[1,5]{Phan Nguyen}
\affil[5]{Department of Thoracic Medicine, Royal Adelaide Hospital, North Terrace, Adelaide, Australia}

\author[1,5]{Arash Badiei}
\author[4]{Nina Eikelis}
\author[4]{Kristopher Nilsen}
\author[4]{Piraveen Pirakalathanan}
\author[1,2,3]{David Parsons}
\author[1,2,3]{Martin Donnelley}

\affil[*]{Corresponding Author: ronan.smith@adelaide.edu.au}

\begin{document}
\maketitle

\begin{abstract}
\textit{Objective:} Endobronchial Valves are a minimally invasive treatment for emphysema. After bronchoscopic placement the valves reduce the flow of air into targeted areas of the lung, causing collapse, and allowing the remainder of the lung to function more effectively. 

\textit{Approach:} X-ray Velocimetry is a novel method that uses X-ray images taken during a breath to track lung motion, producing 3D maps of local ventilation. Healthy sheep received a CT scan and underwent X-ray Velocimetry imaging before and after endobronchial valves were placed in the lung. Sheep were imaged again when the endobronchial valves were removed after 14 days. 

\textit{Main results:} X-ray Velocimetry enabled visualisation and quantification of a reduction of airflow to the areas downstream of the endobronchial valves, both in areas where collapse was and was not visible in CT. Changes to ventilation were also clearly visible in the remainder of the lungs. 

\textit{Significance:} This preclinical study has shown X-ray Velocimetry is capable of detecting changes to ventilation caused by endobronchial valve placement, paving the way towards use in patients.

\end{abstract}


\section*{Introduction}

Endobronchial valves (EBV) can be used to treat gas trapping caused by emphysema \citep{Sciurba2010}. They are one-way valves that allow air to escape from, but not enter, particularly emphysematous areas of the lung, allowing the remainder of the lung to function more effectively. EBVs are inserted with a bronchoscope in a minimally invasive procedure \citep{Klooster2021}. EBV placement has been shown to relieve symptoms \citep{Davey2015, Valipour2016}, with studies showing improvements lasting beyond 6 years \citep{Brown2024}, and they are removable if the treatment is not effective \citep{Gompelmann2018}. EBVs are only effective when placed in areas without collateral ventilation between the target and adjacent lobes \citep{Slebos2017}. Prior to EBV insertion, the Bonchoscopic Chartis Pulmonary Assessment System is used to asses potential sites for collateral ventilation \citep{Herth2013}. The Chartis system allows physiological testing and confirmation of collateral ventilation by blocking lobar airways with a balloon. Once balloon occlusion is confirmed air is allowed to escape the lobe via a central catheter, but air cannot re-enter. If collateral ventilation is present, the lobe does not collapse and airflow is maintained. If collateral ventilation is absent, airflow gradually decreases as the lobe collapses.

Patients treated with EBVs experience improved breathlessness, improved quality of life and improvement in lung function tests. A key radiological feature of successful EBV placement is collapse (atelectasis) that can be visualised using computed tomography (CT) \citep{Slebos2017, Herth2013}.
However, non-invasive measurement of the regional and local changes in airflow caused by valve placement could provide a more accurate and detailed assessment of the relevance and effectiveness of EBV procedures. The first non-invasive localised assessment of their effect on airflow was reported by Salito et.al (2009)  \citep{Salito2009}. They measured the effect of EBVs in excised pig lungs using hyperpolarised helium magnetic resonance imaging \citep{Salito2009}. More recently, Torsani et.al. (2024) have used electrical impedance tomography to track changes in piglets \citep{Torsani2024}. 

Here, we present findings from a pilot study of the effect of EBV placement on regional lung ventilation in healthy sheep using X-ray Velocimetry (XV) functional lung imaging. XV uses X-rays to track the motion of the lung during a breath, creating a 3D map showing how small regions of the lung expand and contract during the breath cycle \citep{Vliegenthart2022, Karmali2023}. XV technology is applicable across a range of lung sizes and species, from mice \citep{Asosingh2022} and rats \citep{Reyne2024} to human adults \citep{Kirkness2023} and children \citep{Bruorton2024}. XV has been shown to correlate to spirometry in healthy patients \citep{Kirkness2023, Siddharthan2023}. Clinical trials have shown promise in looking at altered ventilation in chronic conditions such as burn pit exposure in veterans \citep{Trembley2024}, as well as in interventional studies such as whole lung lavage for silicosis treatment \citep{Barnes2024}. In their trial scanning healthy adults, Siddharthan et al. (2023) showed XV scans had a low median effective radiation dose ranging between 0.41-0.84 mSv \citep{Siddharthan2023}. The ability of XV to localise and quantify ventilation changes has been demonstrated in preclinical rat studies when agar beads were used to intentionally disrupt ventilation within either the left or right lung \citep{Reyne2024_beads}.

The primary aim of this pilot study was to examine the feasibility of using XV to assess the ventilation changes within the lung following EBV insertion. Healthy adult sheep were employed as their lung size is similar to humans. The XV data acquisition protocol for large animals and humans is identical, with fluoroscopic videos taken of individual breaths at 5 angles around the lung, with lobe positioning provided by an accompanying breath-hold CT. These two datasets are processed using the XV Lung Ventilation Analysis Software (XV-LVAS, 4DMedical, Australia) \citep{xv_lvas} to produce respiratory cycle ventilation maps. 

\section*{Materials and Methods}

\subsection*{Animals}
All animal experiments were approved by the South Australian Health and Medical Research Institute (SAHMRI, Adelaide, Australia) Animal Ethics Committee under protocol SAM-22-015. Two male Merino sheep, (weight 63.5 kg and 58 kg) were sedated with a mix of 0.5 mg/kg diazepam and 5 mg/kg ketamine, intubated with an endotracheal tube, and anaesthesia was maintained using inhaled 2-4\% isoflurane. Sheep were ventilated at a tidal volume of 700 ml at 16 breaths/min with an I:E ratio of 1:2, using a maximum pressure of 18 cmH\textsubscript{2}O and a positive end-expiratory pressure of 4 cmH\textsubscript{2}O.   

\subsection*{Imaging}
Breath-hold CT scans were acquired with a Siemens Biograph64 PET/CT scanner. XV scanning was undertaken using a Siemens Artis Zee fluoroscope, using a standard XV acquisition protocol; videos of a ventilated breath were recorded from five different angles around the lungs (AP, ±36°, ±72°). Using this data and the CT scan, ventilation was calculated using XV-LVAS software (4D Medical, Melbourne, Australia) \citep{xv_lvas}.

XV-LVAS uses cross-correlations to track motion in the fluoroscopy videos alongside data from a CT scan to measure the expansion and contraction of the lung during a breath. The output is a 3D map of specific ventilation (SV) values at approximately 5000 voxels across the lung. SV is the change in volume of a voxel of lung during a breath, divided by the starting volume of the voxel. From the SV map, the following parameters can be determined (typically these parameters are calculated across the whole lung, but can also be calculated for smaller subsets from the volumetric XV data): 
\begin{itemize}
    \item Mean Specific Ventilation (MSV): the mean of all specific ventilation values across a given region of the lung
    \item Ventilation Heterogeneity (VH): A measure of how heterogeneous a region of lung is, defined as the interquartile range of specific ventilation values, divided by the MSV of that region. 
\end{itemize}

\section*{Experiment Protocol}
Both sheep were anaesthetised and underwent a breath-hold CT scan followed by an XV scan prior to EBV insertion. Zephyr EBVs (Pulmonx, USA/Switzerland) were then inserted. The first sheep received one standard EBV to the left lung, while the second two EBVs in the right lung After the EBV were inserted, CT and XV scans were repeated and then the animals were allowed to recover and return to the field. Fourteen days later, sheep were anaesthetised as described above and the EBV were removed, with CT and XV scans performed immediately before, and immediately after valve removal. After imaging the sheep were returned to the field for use in unrelated experiments. 

\subsection*{Data Segmentation}
The location of each EBV was identified in the CT scans taken immediately after insertion, and these locations were marked in the pre-treatment CT scans. All areas downstream of these locations were manually segmented from the pre-treatment CT scans by following the airways below the EBV to determine which areas of tissue they serve. A boundary zone approximately \SI{28}{\milli \meter} thick along the edge of the downstream region was separated from the rest of the lung. The left and right lungs were also segmented. These segmentation maps were used for location-based analysis of the XV data, with the XV parameters described above calculated for these regions. 

\section*{Results and Discussion}

\subsection*{Sheep A}

\begin{figure}[h]
    \centering
    \includegraphics[width=0.85\linewidth]{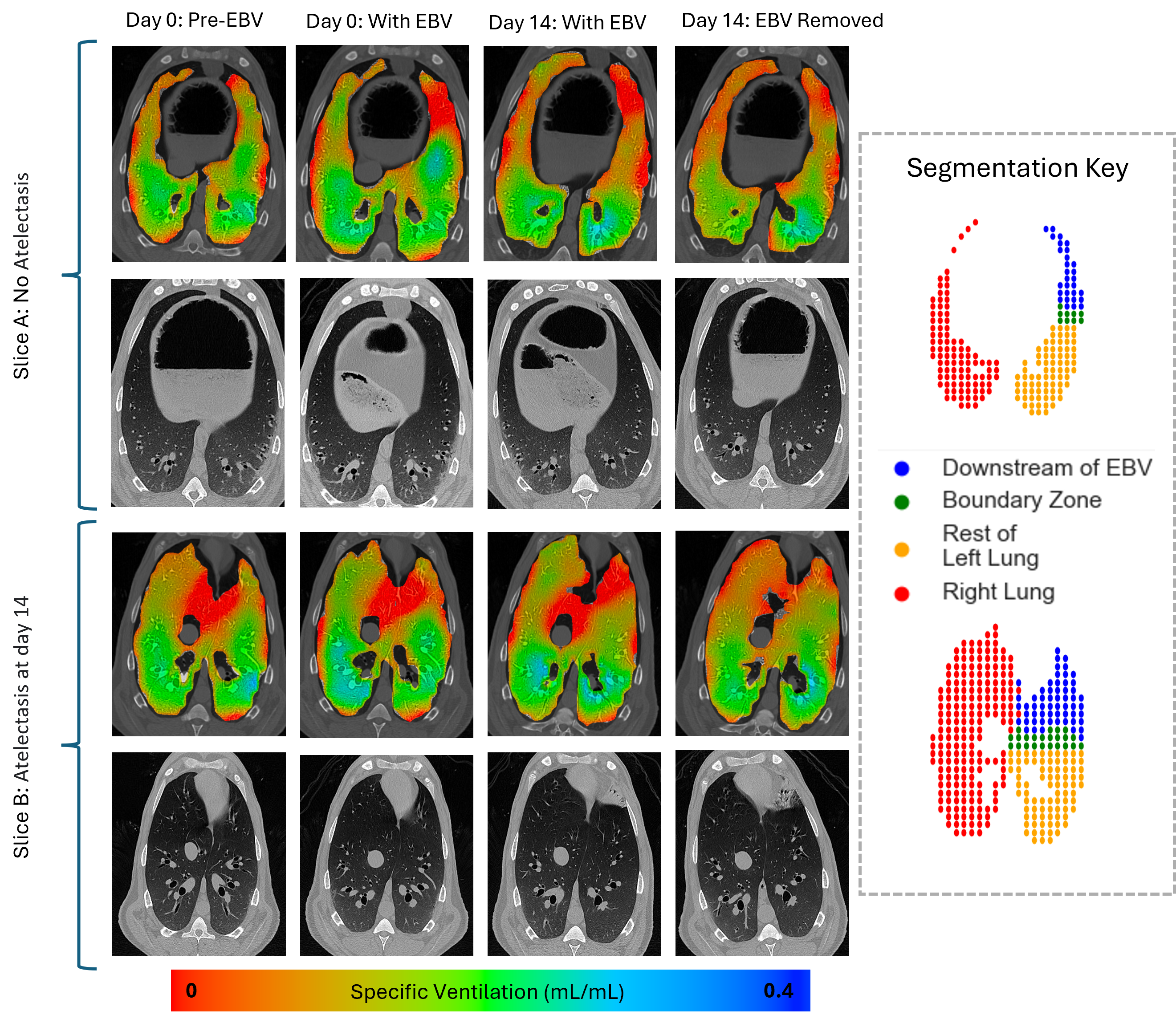}
    \caption{\textbf{Sheep A:} Top row shows XV detects reduced ventilation in the left lung downstream of the EBV after placement, along with corresponding CT slices from this area. Bottom half of figure shows another slice where atelectasis can be seen in part of the downstream region in CT at day 14.}
    \label{fig:md02}
\end{figure}

A single EBV was placed into the left lung. Slice A shown in top row in Figure \ref{fig:md02} shows an immediate reduction in ventilation in the region downstream of the EBV when the valve was placed. Slice B shown in the lower half of the figure shows a slight increase in ventilation in the right lung. There were no clear changes visible in CT immediately after EBV placement. 

After 14 days, atelectasis partially covered the region downstream of the EBV. This was visible in CT slice B, partially filling the region downstream of the EBV, but not in CT slice A. In contrast, XV showed that ventilation was reduced  downstream of the EBV in both slices. 
At day 14 the area of increased ventilation in the right lung was still visible in XV slice B. 

Immediately after the EBV removal, XV showed a slight increase in ventilation in both slices. The region of increased ventilation in slice B appeared to reduce slightly.

\subsection*{Sheep B}

\begin{figure} [h]
    \centering
    \includegraphics[width=0.75\linewidth]{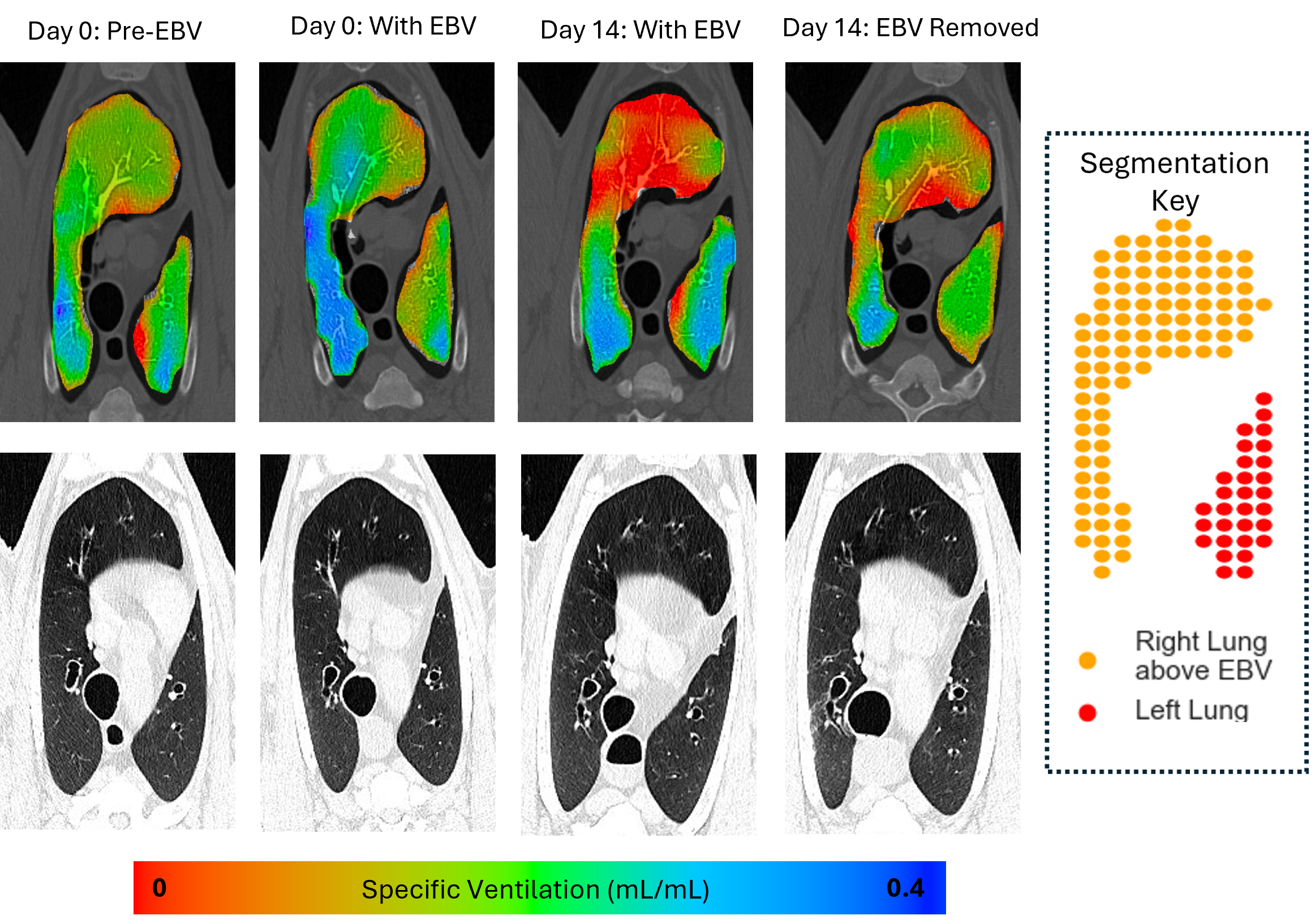}
    \caption{\textbf{Sheep B:} Slices from XV and CT imaging taken from an area of lung above the EBV.  XV shows clear changes in ventilation.}
    \label{fig:md03}
\end{figure}

The second sheep received two EBVs in the right lung. The CT and XV slice shown in Figure  \ref{fig:md03} does not contain any regions directly downstream of the EBV, and is taken from higher up the airway tree than the EBV was placed. The XV slices show a clear change in ventilation immediately after the EBV was inserted, with a large area of high ventilation forming in the right lung. 

At day 14, there was still some areas of increased ventilation, but a large area of reduced ventilation is also visible at the front of the right lung. This does not support the expectation that the EBVs would increase ventilation outside of the downstream region, however an area of increased ventilation was visible in the left lung at this time point. Immediately after removal, the volume of the ventilation defect decreased and the increased ventilation partially returned towards normal. 

We note that the slices shown for both sheep in Figures \ref{fig:md02} and \ref{fig:md03} only show a small subset of the data, and the results shown in the slices may not be representative. Instead, looking at trends in ventilation-based metrics gives a better overall view of the effects in different regions of the lung.  

\subsection*{Quantitative Analysis: Downstream of the EBVs}

The 3D CT volumes and XV ventilation maps were segmented into four regions in which different changes in ventilation were expected to occur with EBV placement and removal (right hand side of Figs. \ref{fig:md02} and \ref{fig:md03}). The region directly downstream of the EBV is where airflow should be blocked, or reduced. This can be seen in figure \ref{fig:quant} as the blue line, with an immediate decrease in MSV in this region in both sheep as soon as the EBVs were added. During the 14 days of the experiment, MSV remained reduced in this region, before increasing in both sheep (to approximately the pre-EBV values) as soon as the EBVs were removed. 

The behavior of the lung and disruption to airflow was not uniform in this region, and in Figure \ref{fig:quant} heterogeneity can be seen to increase as soon as the EBVs were added. This non-uniform behavior is evidenced in the CT images in Figure \ref{fig:md02}, where part of the region collapses, but part remains inflated. VH remained constant in sheep A during the 14 days the EBV was present, while sheep B had a reduction. We suspect that this was likely  due to variability due to the differing EBV placement locations. Upon EBV removal, the heterogeneity downstream of the EBV reduced as airflow was restored to that region. 

\subsection*{Quantitative Analysis: Remainder of Lung}

\begin{figure}
    \centering
    \includegraphics[width=0.6\linewidth]{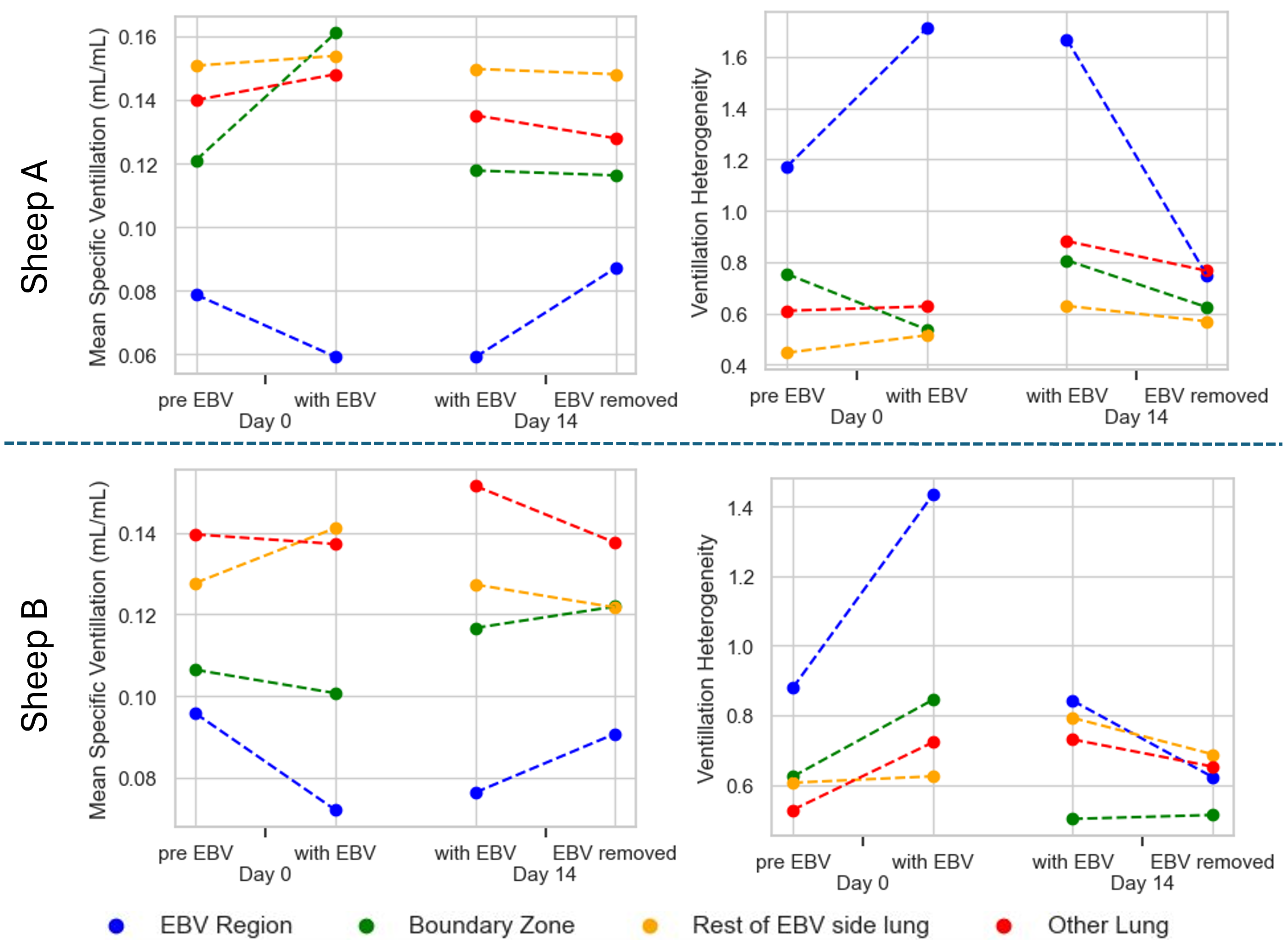}
    \caption{ Changes in Mean Specific Ventilation (MSV - Left) and and Ventilation Heterogeneity (VH - Right) during the course of the experiment for Sheep A (top) and B (bottom). }
    \label{fig:quant}
\end{figure}

The reduction in lung volume in the regions downstream of the EBV should theoretically lead to increased ventilation in the non-affected regions of the lung. Immediately surrounding the downstream-of-EBV region we placed a boundary zone that should have a large increase in ventilation, but may also be subject to errors (both in the segmentation due to changes in lung shape, and due to the sharp change in ventilation being poorly handled by the XV algorithm, as the algorithm has some smoothing applied). Figure \ref{fig:quant} shows that this region behaved differently in the two sheep, with sheep A having an increase in MSV upon insertion, while sheep B had a slight decrease in ventilation. We attribute this decrease to the aforementioned potential errors. 

The remainder of the side of the lung which had the EBV should also see an increase in MSV when the EBV is placed, as can be seen in the yellow lines in Figure \ref{fig:quant}. A similar trend would be expected for the other lung, with sheep A following this expectation. However, sheep B showed a slight reduction in MSV when the EBVs were inserted. This deviation from the expected results is most likely due to the main limitation of our study; these are healthy sheep, and their lungs are not as compliant as emphysematous lungs. 

This limitation of the study is likely the reason there are significant, unexpected changes in the MSV in all regions not downstream of the EBV during the 14 days the EBVs were in place, and during the removal of the EBVs. During this time the lungs were remodeling, leading to changes in airflow. For example, in sheep B, Figure \ref{fig:md03} shows a region of reduced airflow at the front of the lungs appearing at day 14 with the EBV present, despite this area not being downstream of the EBV. This was not expected, however it does explain the reduction in MSV across the region, and also explains why the heterogeneity of the lung has increased.  

\subsection*{Study Limitations and Future Directions}
The main limitation of our study is that the sheep used were healthy, and so do not well model the lungs of the emphysema patients that these EBV are designed to be used in. The way a healthy lung adapts to an EBV will be very different to an emphysematous lung, and so the expected changes to ventilation in the region outside the EBV are not clear. Another limitation of this study is that the sheep were ventilated during their XV scans, and so the mechanics of the breath was slightly different to a free-breathing human. Additionally, the ventilator used was volume driven (i.e. delivering the same volume of air on every breath), and was not adjusted to account for the change in lung volume due to the EBV placement. 

The 14 day period of this study was too short for any long term effects of the EBV to become visible, however the aim of this study was only to show that the effects could be seen in XV imaging. As this is only a pilot study aiming to only show changes can be seen, Chartis Pulmonary Assessment system was not used to assess if there was any collateral ventilation in the regions blocked by the EBVs. Collateral ventilation could explain why a total collapse was not seen in either sheep. 

Rather than continuing this study in an emphysema sheep model (see review by Fehrenbach \citep{Fehrenbach2006}), we believe the results presented show promise for moving towards imaging emphysema patients undergoing EBV placement. This would alleviate the aforementioned limitations. The XV protocol for humans involves free-breathing rather than ventilation \citep{Karmali2023}, alleviating the limitations the ventilator placed on this study. The typical radiation doses of 0.41 - \SI{0.84}{\milli \sievert} as previously described by Siddharthan et.al \citep{Siddharthan2023} are well within the safe limits, allowing for repeated imaging and longitudinal studies. 

This study used qualitative image analysis, alongside simple manual segmentation to give simple quantitative metrics. For the technique to become clinically relevant, we see the need for more advanced image analysis techniques, likely specifically tailored to EBV insertion. For human lungs, tools exist for automated segmentation \citep{Hofmanninger2020}, however tools which provide more meaningful metrics from the XV data are also needed. Additionally, the accuracy and reliability of XV imaging in areas of atelectasis needs to be further investigated. 

\section*{Conclusion}

This study has shown that X-ray Velocimetry is capable of detecting changes to ventilation in the lungs after EBV placement, something which is not possible using standard CT. This study was limited by placing EBVs, devices designed for emphysema treatment, into the lungs of healthy sheep. The non-invasive nature and safe dose rates of XV imaging mean the technique shows promise for using this technique to image patients undergoing EBV placement. 

\section*{Acknowledgement}
The authors acknowledge the facilities and scientific and technical assistance of the National Imaging Facility, a National Collaborative Research Infrastructure Strategy (NCRIS) capability, at the Large Animal Research and Imaging Facility, South Australian Health and Medical Research Institute. We thank Georgia Williams for operating the CT scanner and fluoroscope, Dr Chris Christou for assistance with the animal work, and Ben Maliszewski from Pulmonx for assisting with the EBV selection and preparation.

Studies were supported by the Medical Research Future Fund Grant RFRHPSI000013.

\section*{Ethical Statement}
All animal experiments were approved by the South Australian Health and Medical Research Institute (SAHMRI) animal ethics committee
under protocol SAM-22-015. All relevant sections of the ARRIVE guidelines were followed. 

\section*{Data Availability}
A copy of the data can be found at DOI:10.25909/28191722

\section*{Author Contributions}
Conceptualisation: PN, AB, DP, MD. 
Data Curation: RS, CT, NE, KN, PP.
Formal Analysis: RS, CT, KN.
Funding Acquisition: DP, MD.
Investigation: PN, AB, DP, MD. 
Visualisation: RS.
Writing - Original Draft: RS.
Writing - Review \& Editing: All Authors.

 \bibliographystyle{ieeetr} 
 \bibliography{sample, manual}

\end{document}